\newcommand\sps{\space\space\space\space}
  \def\selectedoptions{final}
\documentclass[
   \selectedoptions
  ]
  {aipproc}

  \def\selectedlayoutstyle{6x9}
\layoutstyle\selectedlayoutstyle

\SetInternalRegister\hbadness{8000} % pseudo latin isn't breaking very well :-)

\newcommand\doingARLO[2][]{%
  \ifx\mmref\undefined #1\else #2\fi
}

\begin{document}

\title
      [Infandum, regina, iubes renovare dolorem]
      {Detrended Fluctuation analysis of Bach's Inventions and Sinfonias pitches
}



\iftrue
\author{G. R. Jafari, P. Pedram and K. Ghafoori Tabrizi}{
  address={Department of physics, Shahid Beheshti University,
Evin, Tehran 19839, Iran \\ Department of nano-science, IPM, P. O.
Box 19395-5531, Tehran, Iran },
 }

\fi

% \copyrightholder{Acoustical Society of America}
\copyrightyear  {2001}

\begin{abstract}
Detrended Fluctuation Analysis (DFA), suitable for the analysis of
nonstationary time series, is used to investigate power law in some
of the Bach's pitches series. Using DFA method, which also is a
well-established method for the detection of long-range
correlations, frequency series of Bach's pitches have been analyzed.
In this view we find same Hurts exponents in the range ($0.7-0.8$)
in his Inventions and sinfonia.

\end{abstract}

\date{\today}

\maketitle

\section{Introduction}
Mathematics and music have some vague sort of affinity, but most
often than not the supposed relationship between the two fields
turns out to consist of complicated structure. The relationship
between mathematics and musical works has been hidden from the
listener since the old days. Thus one is forced to make use of
interpretative techniques in order to search for them, which is
problematic from a methodological point of view. In addition to
mathematics being seen as numerical symbolism, music is closely
linked to absolute physical entities, such as frequency and relation
between intervals (an interval is a space between two notes). It is
an illustrated fact that not just musical notation, but also the
relationship between music and time has something to do with
mathematics. Among great variety of complex and disordered systems
most of  music parameters such as frequency and pitch (pitch is the
sound frequency of any given note) \cite{Gonzalez, lyan, Heather},
Amplitude or Dynamics (dynamics are the changes in volume during a
musical piece) \cite{Jean}, intervals (intervals are the distances
between notes in the musical scale), Rhythm (rhythm is the structure
of the placement of notes in musical time) can be considered as
stochastic processes. Also, some authors try to cluster the music
\cite{Rudi}.

In this paper we characterize the complex behavior of note
frequencies of a selection of Bach's Inventions and Sinfonias
through computation of the signal parameters and scaling exponents.
Inventions and Sinfonias are a collection of short pieces which Bach
wrote for musical education of his young pupils. In music, an
Invention is a short composition (usually for a keyboard instrument)
with two-part counterpoint, which is a broad organizational feature
of much music, involving the simultaneous sounding of separate
musical lines. The Inventions and Sinfonias are two and three voices
music pieces respectively. These voices have independent
characteristic behavior and for simplicity we consider only the
upper voice. Because of non-stationary nature of music frequency
series, and due to finiteness of available data samples, we should
apply methods which are insensitive to non-stationarities, like
trends. In order to separate trends from correlations we need to
eliminate trends in our frequency data. Several methods are used
effectively for this purpose: Detrended Fluctuation Analysis (DFA)
\cite{Peng94}, Rescaled range analysis (R/S) \cite{hurst65} and
Wavelet Techniques (WT) \cite{wtmm}.

%@@@@@@@@@@@@@@@@@@@@@@@@@@@@@@@@
\begin{figure}

\includegraphics[width=11cm]{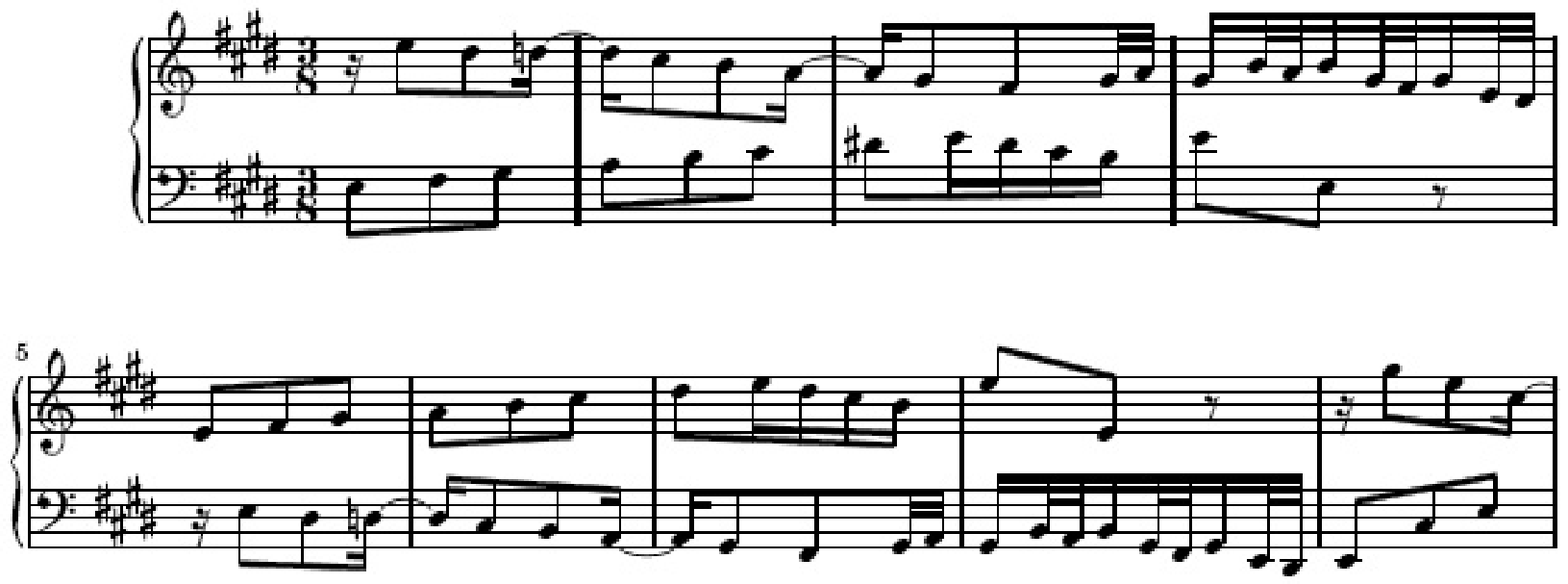}\\
\end{figure}

\begin{figure}
\includegraphics[width=9cm]{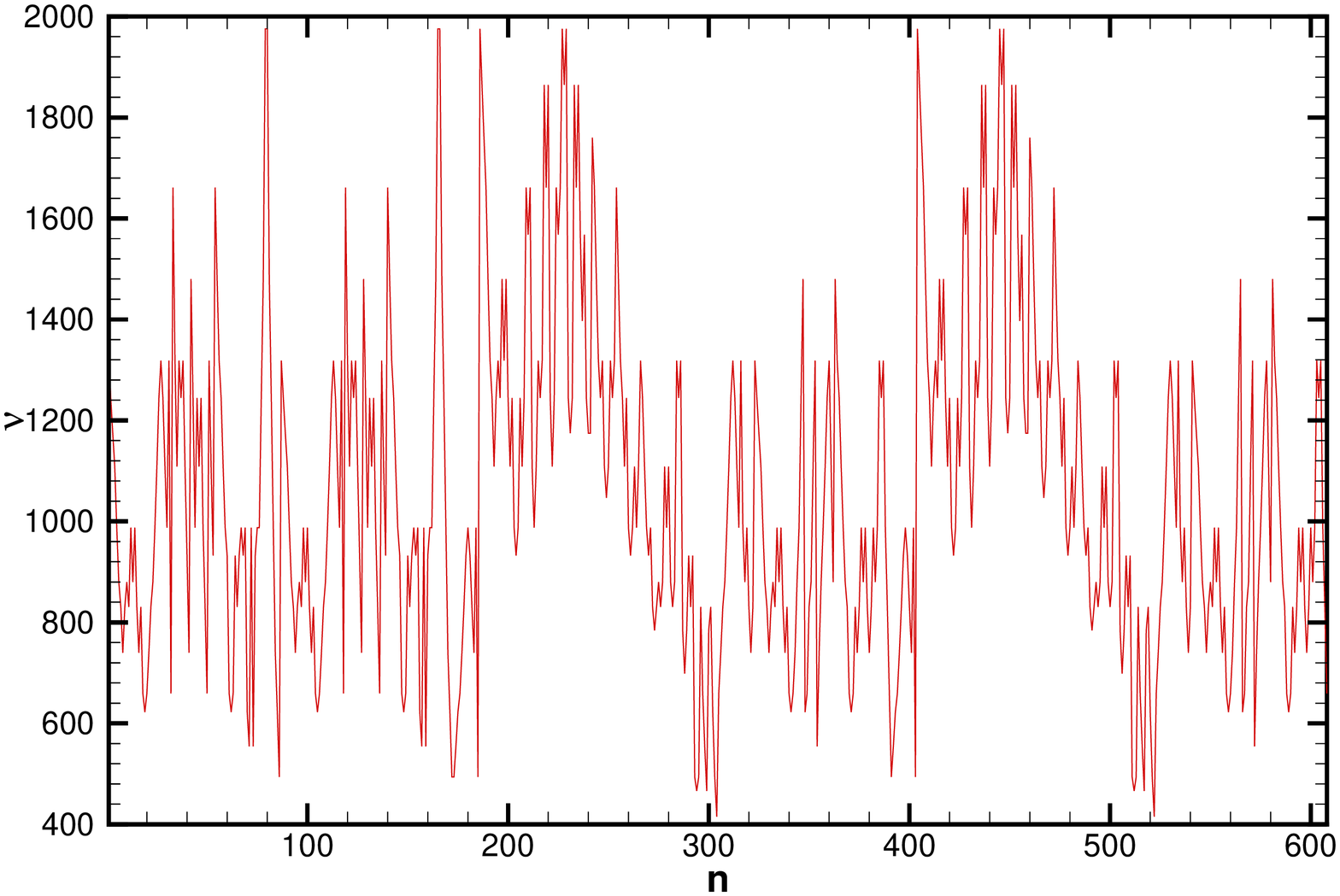}

\caption{Typical up) sheet music and down) frequency series of
Invention No.6 by Bach.} \label{fig1}
\end{figure}
%@@@@@@@@@@@@@@@@@@@@@@@@@@@@@@@@

We use DFA method for the analysis and elimination of trends from
data sets. DFA method introduced by Peng et al. \cite{Peng94} has
became a widely used technique for the determination of (mono-)
fractal scaling properties and the detection of long-range
correlations in noisy, non-stationary time series
\cite{physa,kunhu}. It has successfully been applied to diverse
fields such as DNA sequences \cite{Peng94,DNA}, cardiac dynamics
\cite{cardiac}, climate \cite{climate}, neural receptors
\cite{nural}, economical time series \cite{economics} {\it etc}.

The paper is organized as follows: In section II we describe DFA
methods in details and analyze the frequency series of the
Inventions and Sinfonias. We end the paper by drawing conclusions.

\section{DFA and analysis of music frequency series }

To implement the DFA, let us suppose we have a time series, $N(i)
(i=1,...,N_{max})$ and determine the profile: $
y(j)=\sum_{i=1}^{j}[N(i)-\langle N \rangle]$. Next we break up
$N(i)$ into $K$ non-overlapping time intervals, $I_{n}$, of equal
size $\tau$ where $n=0,1,...K-1$ and $K$ corresponds to the integer
part of $N_{max}/\tau$. In each box, we fit the integrated time
series by using a polynomial function, $y_{pol}(i)$, which is called
the local trend. We detrend the integrated time series $y(i)$ in
each box, and calculate the detrended fluctuation function:
$Y(i)=y(i)-y_{pol}(i)$. For a given box size $s$, we calculate the
root mean square fluctuation:
\begin{equation}
F(s)=\sqrt{\frac{1}{N_{max}}\sum_{i=1}^{N_{max}}[Y(i)]^{2}}\label{FDFA}.
\end{equation}
The above computation is repeated for box sizes $s$ (different
scales) to provide a relationship between $F(s)$ and $s$. A power
law relation between $F(s)$ and $s$ indicates the presence of
scaling
\begin{equation}
F(s)\sim s^{H}.\label{Fs}
\end{equation}
The parameter $H$, called Hurst exponent, represents the correlation
properties of the signal: if $H=0.5$, there is no correlation and
the signal is an uncorrelated signal \cite{Peng94}; if $H<0.5$, the
signal is anticorrelated; if $H>0.5$, there are positive
correlations in the signal. In the two latest cases, the signal can
be well approximated by the fractional Brownian motion law
\cite{Feder88}. Also, the auto correlation function can be
characterized by a power law $C(s)\equiv\langle N_kN_{k+s} \rangle
\sim s^{-\gamma}$ with $\gamma=2-2H$. Its power spectra can be
characterized by $S(\omega)\sim\omega^{-\beta}$ with frequency
$\omega$ and $\beta=2H-1$. In non-stationary case, correlation
exponent and power spectrum scaling are $\gamma=-2H$ and
$\beta=2H+1$, respectively \cite{Peng94,eke02}.

\begin{figure}[t]
\includegraphics[width=8.5cm]{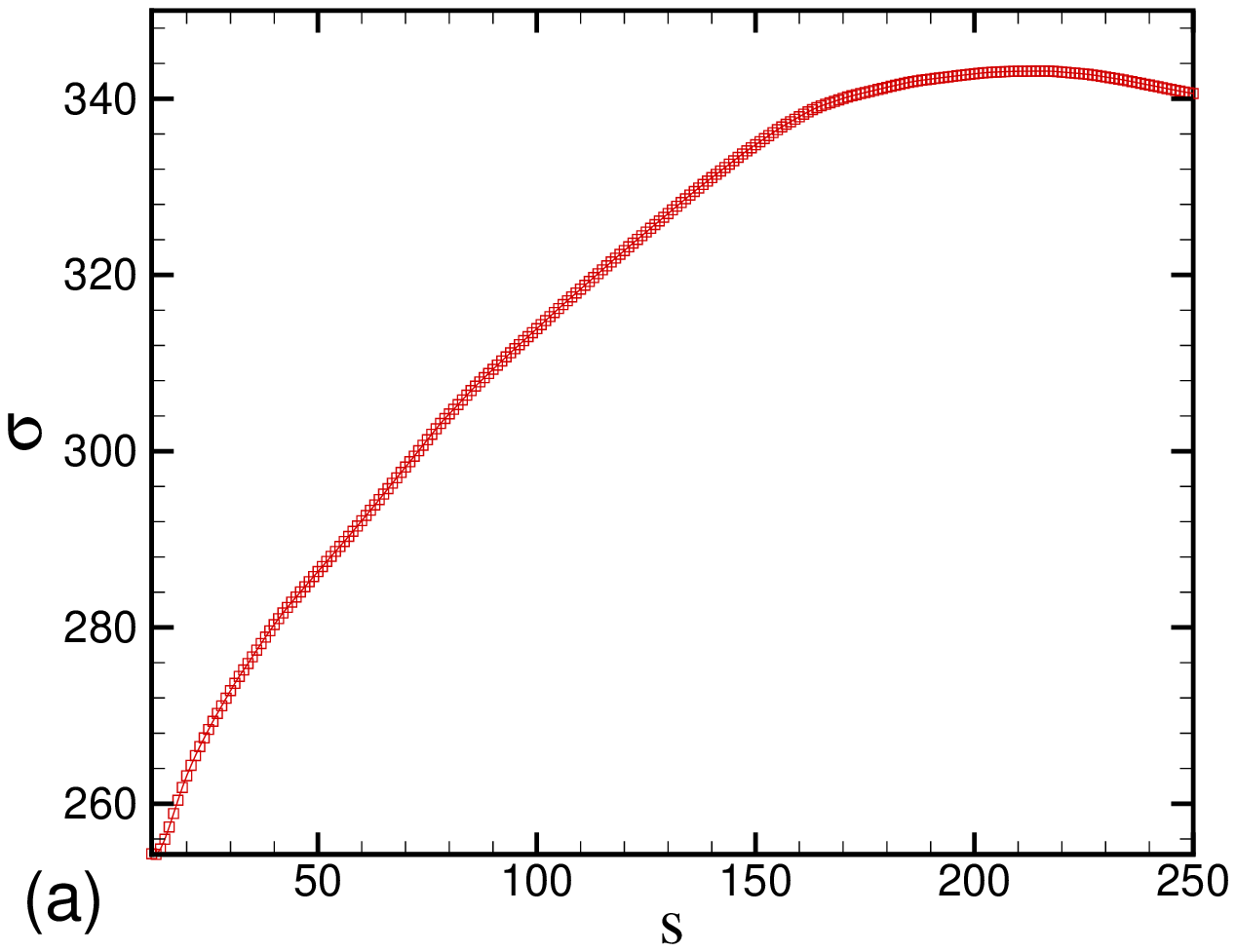}
\includegraphics[width=8.5cm]{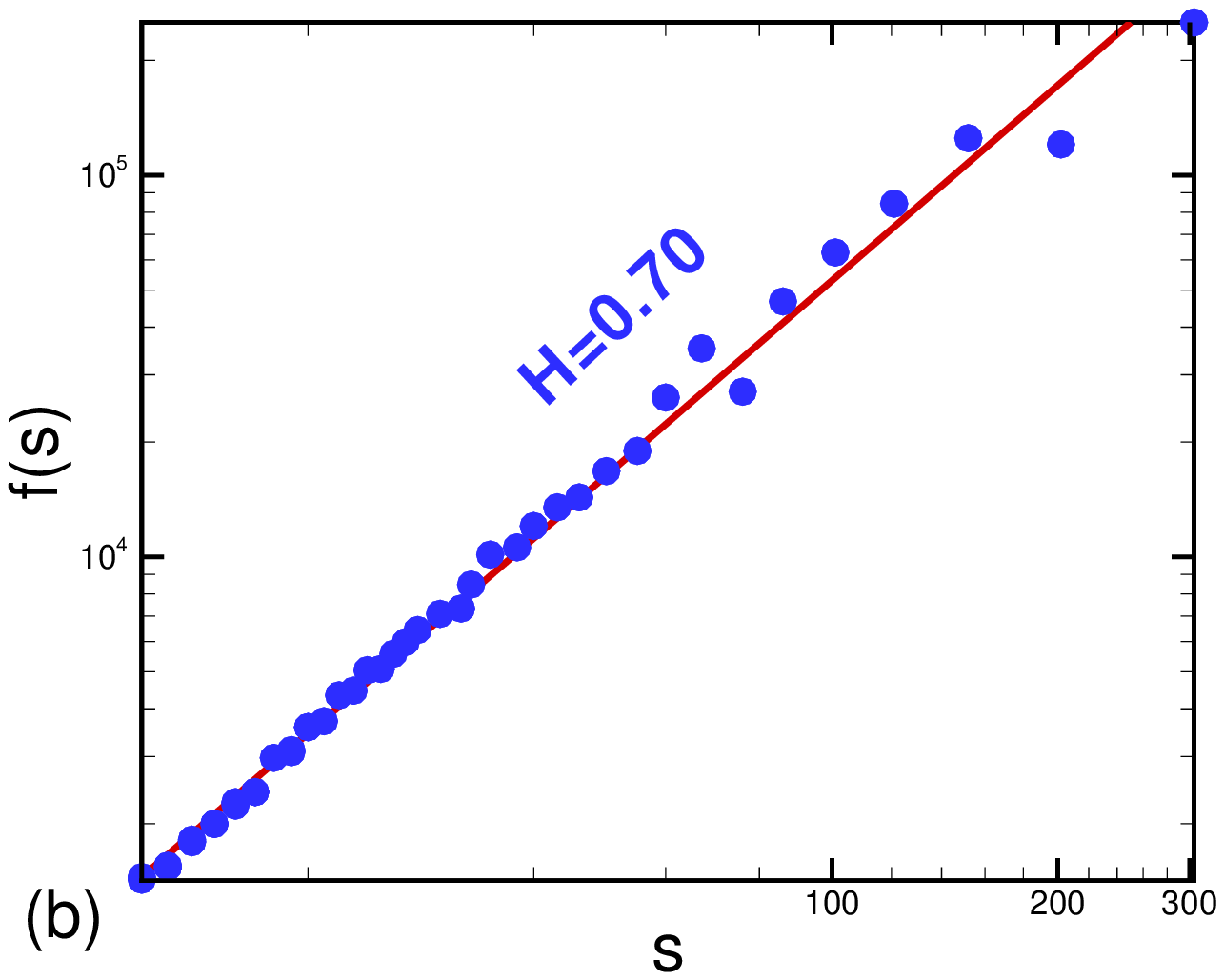}
\caption{\small
  a) $s$ dependence of variation and  b) The log-log plot $F(s)$ versus $s$ for Invention
no.6 frequency series.} \label{fig2}
\end{figure}

It can be checked out that, frequency series is non-stationary. One
can verify non-stationarity properties experimentally by measuring
stability of average and variance in moving windows by, for example,
using scale $s$ (Fig. \ref{fig2}a).
In Fig. \ref{fig2}b we plot in double-logarithmic scale the
corresponding fluctuation function $F(s)$ against the box size s.
Using the above procedure, we obtain the following estimate for the
Hurst exponent: $H=0.70\pm 0.03$. The $F(s)$ exhibits an approximate
scaling regime from $s=2$ up to nearly $s=6$ (in logarithmic scale).
Since $H>0.5$ it is concluded that the frequency series show
persistence; i.e. strong correlations between consecutive
increments. The values which  derived for quantities of DFA1 method
for Invention no. 6 are given in Table \ref{Tab1} (second line). We
have calculated the Hurst exponent for other Inventions and Sinfonia
as well, all being in the $0.7-0.8$ range (Table \ref{Tab1}).

Usually, in DFA method, deviation from a straight line in log-log
plot of Eq.(\ref{Fs}) occurs for small scales $s$. These deviations
are intrinsic to the usual DFA method, since the scaling behavior is
only approached asymptotically. Deviations limit the capability of
DFA to determine the correct correlation behavior in very short
records and in the regime of small $s$. The modified DFA is defined
as follows \cite{physa}:
\begin{eqnarray}
F^{\rm mod}(s) &=&  F(s) {\langle [F^{\rm shuf}(s')]^2 \rangle^{1/2}
\, s^{1/2} \over \langle [F^{\rm shuf}(s)]^2 \rangle^{1/2} \,
s'^{1/2} } \quad {\rm (for} \,\,\, s' \gg 1), \label{fmod}
\end{eqnarray}
where $\langle [F^{\rm shuf}(s)]^2 \rangle^{1/2}$ denotes the usual
DFA fluctuation function, defined in Eq.(\ref{FDFA}), averaged over
several configurations of shuffled data taken from original series,
and $s' \approx N/20$. The improvement is very useful especially for
short records or records that have to be split into shorter parts to
eliminate problematic nonstationarities, since the small $s$ regime
can be included in the fitting range for the fluctuation exponent
$H$.

The modified DFA method indicates the correct correlation behavior
also in presence of broadly distributed data, where the common DFA
fails to distinguish long-range correlations from deviations caused
by broad distributions. The values of Hurst exponent obtained by
modified DFA1 methods for frequency series is $0.72\pm0.03$. The
relative deviation of Hurst exponent which is obtained by modified
DFA1 in comparison to DFA1 for original data is less than $4\%$.
\begin{table}[t]
%\begin{center}
\caption{\label{Tab1} Values of Hurst $(H)$, power spectrum
scaling $(\beta)$ and auto-correlation scaling $(\gamma)$
exponents for the selections of Inventions and sinfonias frequency
series obtained by DFA1.}
%\medskip
\begin{tabular}{|c|@{\hspace{0.3cm}}c@{\hspace{0.3cm}}|@{\hspace{0.3cm}}c@{\hspace{0.3cm}}|@{\hspace{0.3cm}}c@{\hspace{0.3cm}}|}
  \hline
     & $H$ & $\beta$&$\gamma$\\ \hline
Invention no.1 & $0.77\pm 0.03$
&$2.54\pm0.06$&$-1.54\pm0.03$\\\hline

 Invention no.6 & $0.72\pm 0.03$ &$2.44\pm0.06$  &$-1.44\pm0.03$     \\ \hline

Sinfonia no.1 & $0.71\pm 0.03$
&$2.42\pm0.06$&$-1.42\pm0.03$\\\hline

Sinfonia no.13 & $0.73\pm 0.03$
&$2.46\pm0.06$&$-1.46\pm0.03$\\\hline

\end{tabular}
%\end{center}
\end{table}
\section{Conclusion}
DFA is a scaling analysis method used to quantify long-range
power-law correlations in signals. Many physical and biological
signals are `noisy', heterogeneous and exhibit different types of
nonstationarities, which can affect the correlation properties of
these signals. Applying DFA1 method demonstrates that the music
frequency series have long range correlation. We calculated Hurst
exponent for other Inventions and Sinfonia and found it to be in the
$0.7-0.8$ range.

\section{Acknowledgment}
GRJ would like to acknowledge the hospitality extended during his
visit at the IPM, UCLA, where this work was started.

\doingARLO[\bibliographystyle{aipproc}]
          {\ifthenelse{\equal{\AIPcitestyleselect}{num}}
             {\bibliographystyle{arlonum}}
             {\bibliographystyle{arlobib}}
          }

\end{document}